\def\year{2021}\relax
\documentclass[letterpaper]{article} 
\usepackage{amsfonts}
\usepackage{amssymb}
\usepackage{amsthm,amsmath}
\usepackage{mathrsfs}
\usepackage{indentfirst}
\usepackage{multirow}
\usepackage{subfigure}
\usepackage{bm}
\usepackage{graphicx}
\usepackage[title]{appendix}
\usepackage{aaai22}  
\usepackage{times}  
\usepackage{helvet} 
\usepackage{courier}  
\usepackage[hyphens]{url}  
\usepackage{graphicx} 
\usepackage{natbib}  
\usepackage{caption} 
\title{ProSTformer: Pre-trained Progressive Space-Time  Self-attention Model for Traffic Flow Forecasting  }
\author {
    Xiao Yan,\textsuperscript{\rm 1}
    Xianghua Gan, \textsuperscript{\rm 1}
    Jingjing Tang \textsuperscript{\rm 1} 
    Rui Wang \textsuperscript{\rm 1}\\
}

\affiliations {
    \textsuperscript{\rm 1} School of Business Administration, Faculty of Business Administration, Southwestern University of Finance and Economics. \\
}

\begin{document}
\maketitle

\begin{abstract}
Traffic flow forecasting is essential and challenging to intelligent city management and public safety. Recent studies have shown the potential of convolution-free Transformer approach to extract the dynamic dependencies among complex influencing factors. However, two issues prevent the approach from being effectively applied in traffic flow forecasting. First, it ignores the spatiotemporal structure of the traffic flow videos. Second, for a long sequence, it is hard to focus on crucial attention due to the quadratic times dot-product computation. To address the two issues, we first factorize the dependencies and then design a progressive space-time self-attention mechanism named ProSTformer. It has two distinctive characteristics: (1) corresponding to the factorization, the self-attention mechanism progressively focuses on spatial dependence from local to global regions, on temporal dependence from inside to outside fragment (i.e., closeness, period, and trend), and finally on external dependence such as weather, temperature, and day-of-week; (2) by incorporating the spatiotemporal structure into the self-attention mechanism, each block in ProSTformer highlights the unique dependence by aggregating the regions with spatiotemporal positions to significantly decrease the computation. We evaluate ProSTformer on two traffic datasets, and each dataset includes three separate datasets with big, medium, and small scales. Despite the radically different design compared to the convolutional architectures for traffic flow forecasting, ProSTformer performs better or the same  on the big scale datasets than six state-of-the-art baseline methods by RMSE. When pre-trained on the big scale datasets and transferred to the medium and small scale datasets, ProSTformer achieves a significant enhancement and behaves best.

\end{abstract}

\setcounter{secnumdepth}{2}
\section{Introduction}

\begin{figure}[ht] \centering
    \subfigure[Outflows from region $\bigstar$] {
        \label{space:a}
    \includegraphics[width=0.4\columnwidth]{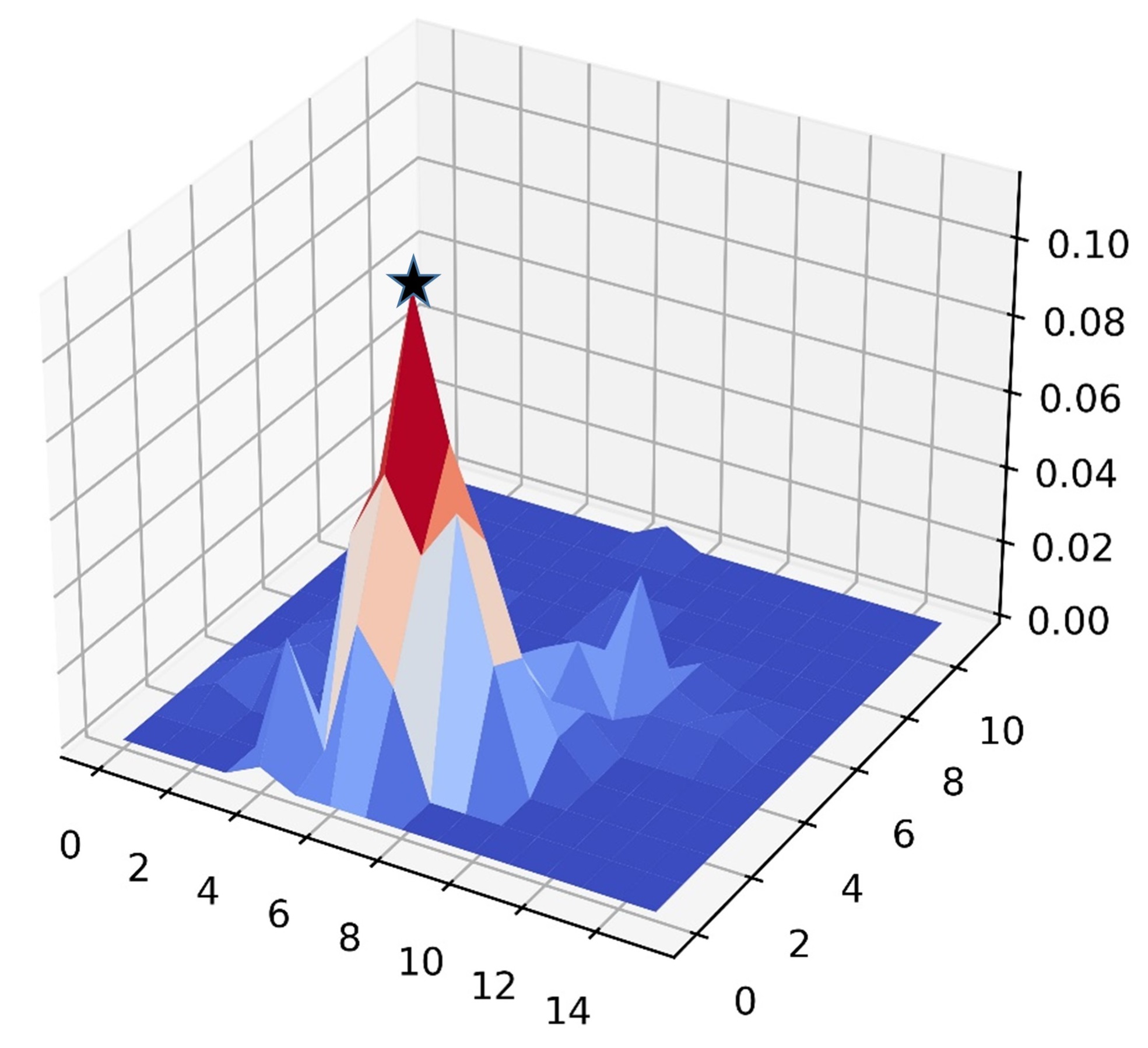}
    }
    \subfigure[Closeness] {
    \label{time:b}
    \includegraphics[width=0.4\columnwidth]{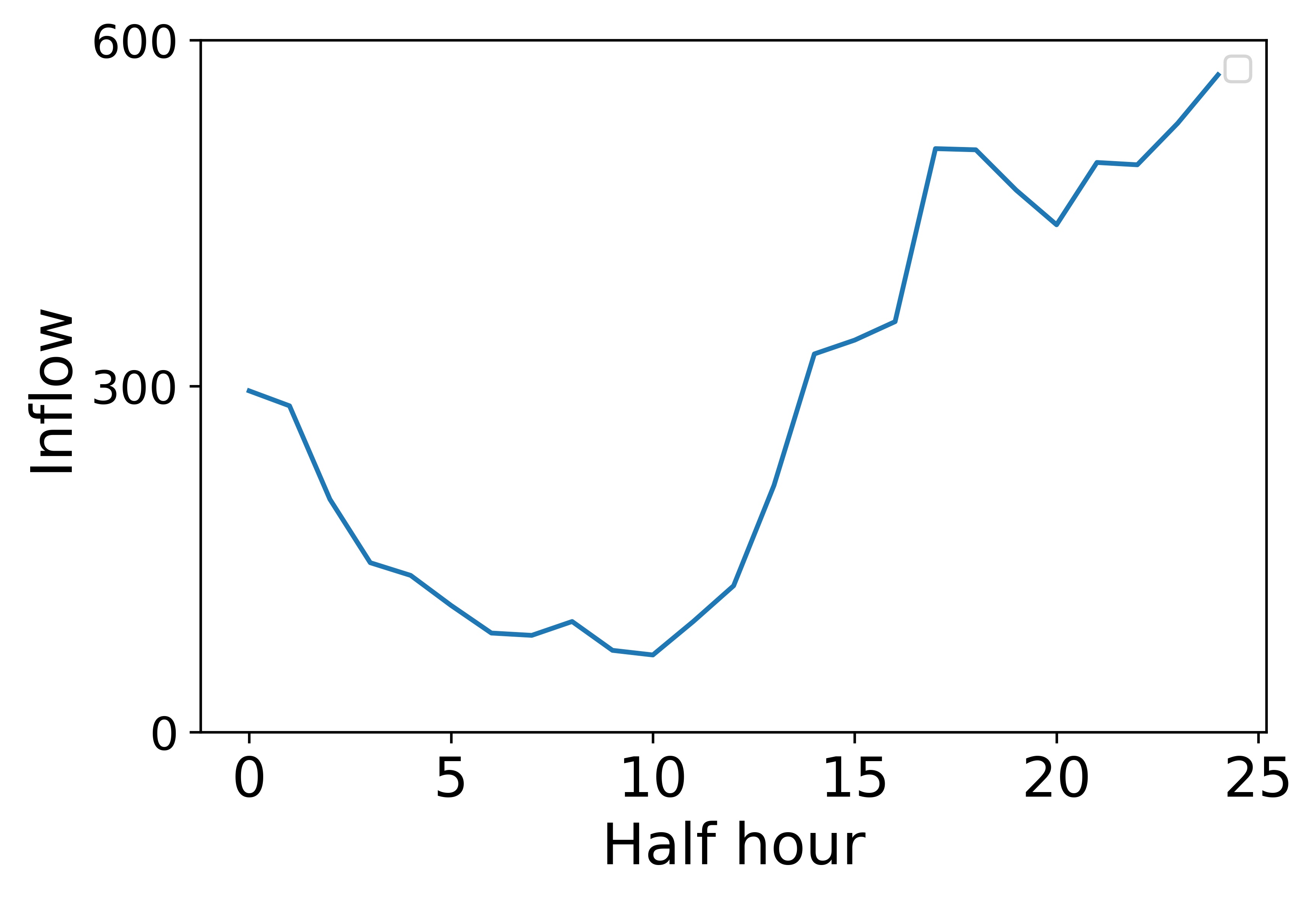}
    }
    \subfigure[Period] {
        \label{time:c}
        \includegraphics[width=0.4\columnwidth]{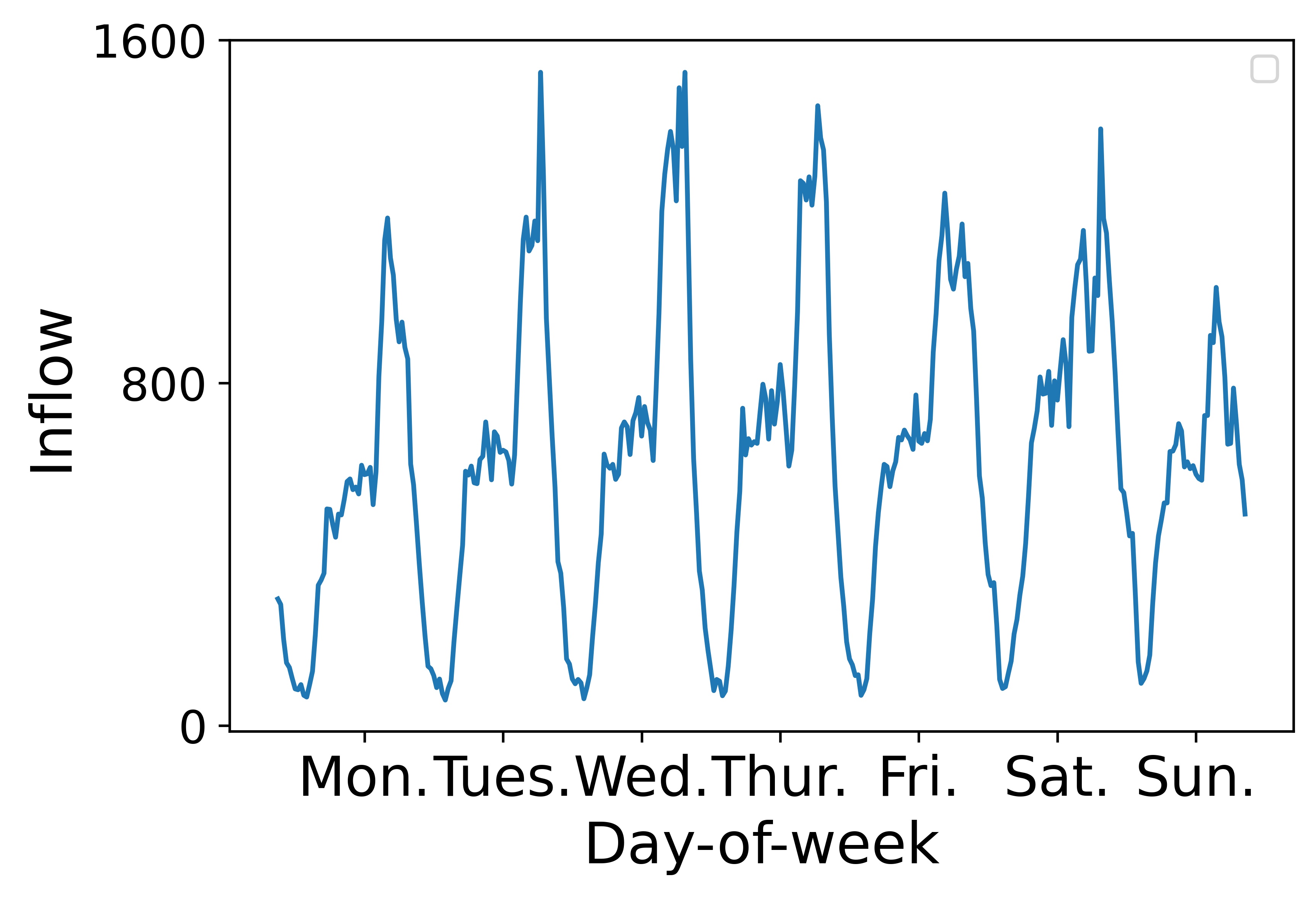}
        }
        \subfigure[Trend] {
    \label{time:d}
    \includegraphics[width=0.4\columnwidth]{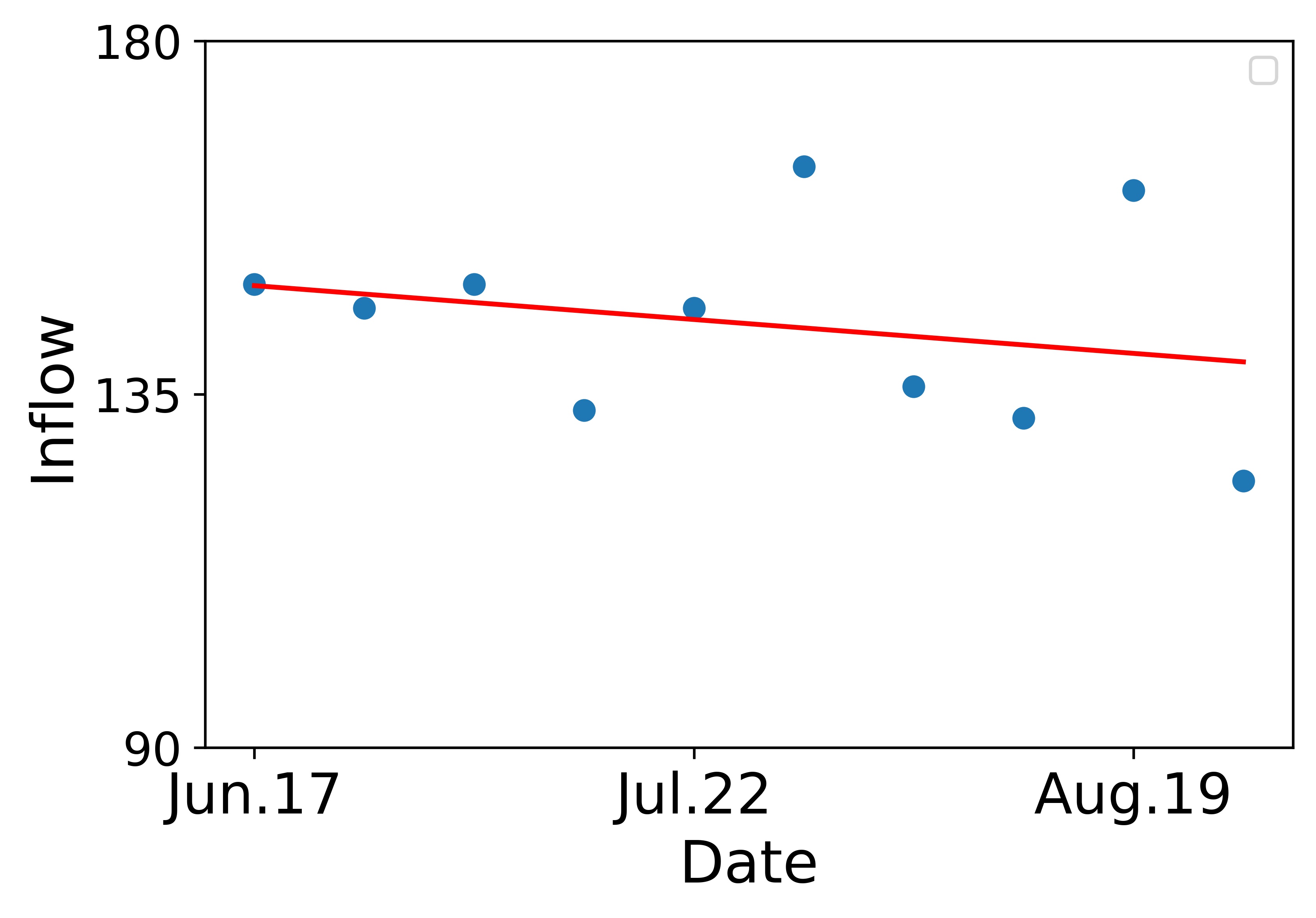}
    }
    \caption{The spatial dependence and temporal dependence. (a) reflects the local and global spatial dependencies. (b), (c), and (d) reflect the temporal dependencies of closeness, period, and trend, specifically, (b) shows that flows are relevant to recent time intervals;
    (c) shows that flows during certain hours are similar on consecutive days; (d) shows that long-term flows progressively decrease from June to August.}
    \label{space-time dependence}
    \end{figure}
Traffic flow forecasting plays a vital role in traffic control, vehicle
scheduling, and risk assessment \citep{zheng2014urban}. For example, it is indispensable for
congestion alleviation and real-time control of traffic signals in intelligent
cities. In bike-sharing systems, bike flow forecasting is crucial for
operators to rebalance bikes from oversupplied regions to undersupplied regions.
It is a standard forecasting approach to generate flow videos for partitioned city regions and then leverage flow videos and external factors such as weather conditions, wind speed, temperature, and day-of-week.
Works using this approach include \citet{zhang2016dnn}, \citet{zhang2017deep}, \citet{Ke2017}, \citet{yao2018modeling}, \citet{zhang2019short}, \citet{du2019deep}, and \citet{chen2021multiple}.
These studies adopt various convolutional networks to extract the dynamic spatiotemporal dependence.
Recently,  \citet{duan2019pre} and \citet{lin2019spatial} not only use various convolutional networks, but also introduce Transformer to construct model architecture for traffic flow forecasting. In these studies, Transformer is applied in conjunction with convolutional networks.
\citet{dosovitskiy2020image} show that the reliance on convolutional networks is not necessary and a pure Transformer network applied directly to sequences of image  patches can effectively extract the spatiotemporal dependence.
When pre-trained on large amounts of data and transferred to multiple mid-sized or small image recognition benchmarks, Transformer performs similar to or better than state-of-the-art convolutional networks.
This study shows the potential of convolution-free Transformer approach for traffic flow forecasting.

However, two significant issues prevent the pure Transformer from being effectively applied in traffic flow forecasting.
First, Transformer does not incorporate spatiotemporal structure of flow videos into self-attention mechanism.
Take NYCTaxi dataset as an example, Figure \ref{space:a} depicts outflows from region
$\bigstar$ to itself and other regions from 6 PM to 7 PM.
The percentages of the outflows from $\bigstar$ to itself,
from $\bigstar$ to adjacent regions (i.e., the surrounding 14
regions),
and from $\bigstar$ to distant regions (i.e., other 28 nonzero regions)
are about 11\%, 64\%, and 25\%, respectively.
Figure \ref{time:b}, \ref{time:c}, and \ref{time:d} show temporal dependencies
of closeness, period, and trend, respectively.
The spatiotemporal structure, which reflects the local and global spatial
dependencies and temporal correlations among different time intervals,
is ignored in Transformer architecture.
Second, by splitting the flow videos into a long sequence of patches, 
a large part of low relevant patches make the traditional self-attention mechanism hard to focus on crucial attention due to the quadratic times dot-product computation.

To address the two issues, we factorize the dependencies into local-global spatial dependence, inside-outside temporal dependence, and external dependence such as weather, temperature, and day-of-week.
Corresponding to the factorization, we propose a progressive space-time self-attention mechanism named ProSTformer.
It progressively focuses on spatial dependence from local to global regions, on temporal dependence from inside to outside fragment (i.e., closeness, period, and trend), and finally on external dependence.
In ProSTformer, each block highlights the unique dependence by aggregating the regions with spatiotemporal positions to significantly decrease the computation.

There are some prior works on improving the efficiency of self-attention to alleviate issue 2. Sparse Transformer \citep{child2019generating}, LogSparse Transformer \citep{li2019enhancing}, Longformer \citep{beltagy2020longformer}, Reformer \citep{kitaev2020reformer}, and  Informer \citep{zhou2021informer} focus on reducing the complexity of the self-attention mechanism.
E.g., Informer designs a ProbSparse self-attention mechanism distilling highlights dominating self-attention by halving cascading layer input and efficiently handles extreme long input sequences.
TimeSformer \citep{bertasius2021space} use a divided time-space self-attention mechanism on video classification tasks to decrease the computation.

The contributions of this paper are summarized as follows:
\begin{itemize}
    \item We propose a progressive space-time self-attention mechanism named ProSTformer that can enhance the prediction capacity in traffic flow forecasting problem.
    We also show the convolution-free Transformer-like model's potential value to capture spatiotemporal dependence in traffic flow forecasting.
    \item ProSTformer incorporates the structure information of input.
    In ProSTformer, each block highlights the unique dependence by aggregating the regions with spatiotemporal positions to significantly decreases the computation.
    \item We conduct data augmentation tasks for ProSTformer, and show that closely relevant and large amounts of data pre-training both are of great importance for traffic flow forecasting.

\end{itemize}

\section{Preliminary}
\begin{figure}[h] \centering
    \subfigure[Outflow matrix by half-hour] {
        \label{fig:a}
    \includegraphics[width=0.27\columnwidth]{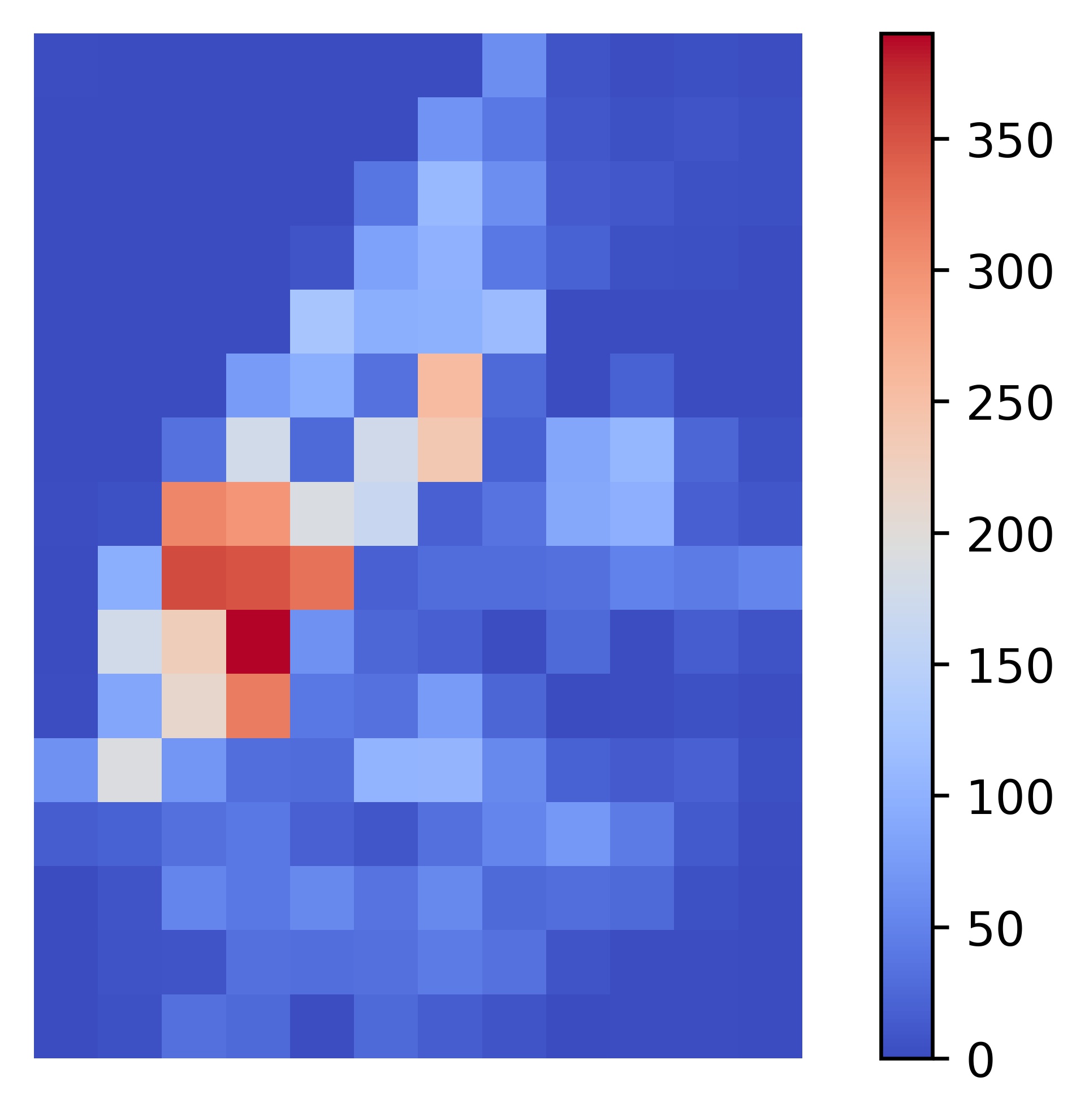}
    }
    \subfigure[Inflow and outflow] {
    \label{fig:b}\hspace{5mm}
    \includegraphics[width=0.3\columnwidth]{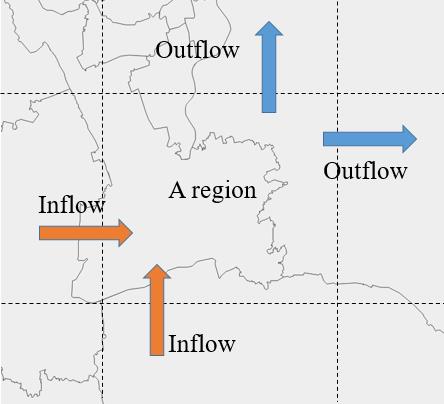}
    }
    \caption{Regions in New York City}
    \label{Regions in New York City}
    \end{figure}

We first define the problem of traffic flow forecasting.
\newtheorem{mydef}{Definition}
\newtheorem{mypro}{Problem}
\begin{mydef}
    \textbf{(Region and time partition \citep{zhang2016dnn})}
    The city area is partitioned into $I \times J$ grids uniformly based on the longitude and latitude where a grid denotes a region, as shown in Figure \ref{Regions in New York City}.
    \end{mydef}

\begin{mydef}
  \textbf{(Inflow/outflow \citep{zhang2016dnn})}
    Let $R$ be a collection of trajectories at the $t^{th}$ time interval. For a grid $(i, j)$ that lies at the $i^{th}$ row and the $j^{th}$ column, the inflows and outflows at time interval $t$ are defined respectively as:
    \end{mydef}
\begin{align*}
    x_{t}^{in, i, j} &=\sum_{Tr \in R} \mid\left\{k \geq 1 \mid  g_{k}^{end} \in(i, j)\right\}  \\
    x_{t}^{out, i, j} &=\sum_{Tr \in R} \mid\left\{k \geq 1 \mid
    g_{k}^{start} \in(i, j) \right\},
    \end{align*}
where $\operatorname{Tr}: g_{1} \rightarrow g_{2} \rightarrow \cdots
\rightarrow g_{|T r|}$ is a trajectory in $R$, and $g_{k}^{start}$,
$g_{k}^{end}$ are the geospatial coordinate; $g_{k}^{start} \in(i, j)$,
$g_{k}^{end} \in(i, j)$ mean the trajectory start or end in the grid $(i, j)$,
note that the trajectory can start and end in the same region,
$| \cdot  |$ denotes the cardinality of a set.

At time interval $t$, inflows and outflows in all $I \times  J$ regions can be
denoted by a tensor $\mathbf{X}_{t} \in R^{2 \times I \times J}$ where $\left(\mathbf{X}_{t}\right)_{0, i, j}=x_{t}^{i n, i, j},\left(\mathbf{X}_{t}\right)_{1, i, j}=x_{t}^{\text {out }, i, j}$.
The outflow matrix is shown in Figure \ref{fig:a}.

\begin{mypro}
    Predict $\mathbf{X}_{n}$ given historical observations $\{ \mathbf{X}_{t}| t=0,\cdots ,n-1 \} $ and external factors such as weather conditions, wind speed, temperature, and day-of-week.
    \end{mypro}

\section{Methodology}
\textbf{Input \citep{zhang2017deep}.}
We sample historical flow videos from recent time to near history and distant history according to three corresponding temporal views: closeness, period, and trend.
We select hours, daily, and weekly as the key timesteps to construct the three views.
For each of temporal views, we fetch a list of key timesteps’ flow matrices and concatenated them, to construct the input as:
\begin{align*}
    &\mathbf{X_{closeness}}=\left[\mathbf{X}_{t-1}, \mathbf{X}_{t-2}, \cdots, \mathbf{X}_{t-l_{r}}\right] \in \mathbb{R}^{N \times C \times l_{r}} \\
    &\mathbf{X_{period}}=\left[\mathbf{X}_{t-p_{d}}, \mathbf{X}_{t-2 p_{d}}, \cdots, \mathbf{X}_{t-l_{d} * p_{d}}\right] \in \mathbb{R}^{N \times C \times l_{d}} \\
    &\mathbf{X_{trend}}=\left[\mathbf{X}_{t-p_{w}}, \mathbf{X}_{t-2 p_{w}}, \cdots, \mathbf{X}_{t-l_{w} * p_{w}}\right] \in \mathbb{R}^{N \times C \times l_{w}} \\
    &\{\mathbf{X}_{t}| t=0,\cdots ,n-1 \}=\left[\mathbf{X_{closeness}} \mathbf{X_{period}}, \mathbf{X_{trend}} \right] ,
    \end{align*}
where  $l_{r},l_{d},l_{w}$ are input lengths of  hours, daily, and weekly, $p_{d}$, $p_{w}$ are daily and weekly periods.
We choose $l_{r},l_{d},l_{w}$ as 4, 4, 4,  in our study.

\textbf{Decomposition into patches.}\label{decom} Similar to
\citet{dosovitskiy2020image}, we first decompose each frame into $N_1$ big patches, each of size $2 \times P_1 \times P_1^{\prime}$ and  2 means outflow and inflow. Then we decompose each big patch into $N_2$ small patches, each of size
$2 \times P_2 \times P_2^{\prime}$. We obtain $N$ patches, where $N=N_1\times
N_2=HW/P_{2}P_2^{\prime}$.
We then flatten these patches of the entire flow videos into vectors
$\mathbf{x}_{(p,t)}\in  \mathbb{R}^{2 \times P_2 \times P_2^{\prime}}$, where
$p=1,\ldots ,N$ denotes spatial localization
and $t=1,\ldots ,F$ denotes indexes of frames in the flow videos. $F=F_1 \times F_2$ includs $F_1$ fragments (i.e., closeness, period, trend), and each fragment include $F_2$ frames, i.e., $F_1=3, F_2=4$ in our study.

\textbf{Linear embedding.} By means of a learnable matrix $E \in  \mathbb{R}^{D \times 2 P_2P_2^{\prime}}$, we linearly map each patch $x_{(p,t)}$ into an
embedding vector $\mathbf{z}_{(p,t)}^{(0)} \in \mathbb{R}^{D}$:
\begin{equation}\label{linea_embeding}
    \mathbf{z}_{(p, t)}^{(0)}=E \mathbf{x}_{(p, t)}+\mathbf{e}_{(p, t)}^{p o s},
\end{equation}
where $\mathbf{e}_{(p, t)}^{p o s} \in \mathbb{R}^{D}$ denotes a learnable positional embedding added to encode the spatiotemporal position of each patch.
The final sequence of embedding vectors  $\mathbf{z}_{(p, t)}^{(0)}$ for $p=1,\ldots,N$ and $t=1,\ldots,F$ represents the input flow videos after preprocessing.

\textbf{Query-Key-Value.} Similar to the BERT Transformer \cite{bert}, ProSTformer also uses Encoder framework consisting of $L$ encoding blocks.
At each block $l$, a query/key/value vector is computed for each patch from the $\mathbf{z}_{(p, t)}^{(l-1)}$ encoded by the preceding block:
\begin{equation}\label{qkv}
    \begin{aligned}
        \mathbf{q}_{(p, t)}^{(l, a)} &=W_{Q}^{(l, a)} \operatorname{LN}\left(\mathbf{z}_{(p, t)}^{(l-1)}\right) \in \mathbb{R}^{D_{h}}\\
        \mathbf{k}_{(p, t)}^{(l, a)} &=W_{K}^{(l, a)} \operatorname{LN}\left(\mathbf{z}_{(p, t)}^{(l-1)}\right) \in \mathbb{R}^{D_{h}}\\
        \mathbf{v}_{(p, t)}^{(l, a)} &=W_{V}^{(l, a)} \operatorname{LN}\left(\mathbf{z}_{(p, t)}^{(l-1)}\right) \in \mathbb{R}^{D_{h}},
        \end{aligned}
\end{equation}
where $LN$ denotes LayerNorm \cite{ba2016layer}, $a=1,\ldots,A$ is an index over multiple self-attention heads, and $A$ denotes the total number of self-attention heads. The latent dimensionality for each self-attention head is set to $D_{h}=D/A$.

\subsection{\emph{Progressive} Space-Time self-attention.}
\begin{figure}[ht]
    \centering
    \includegraphics[width=0.5\textwidth]{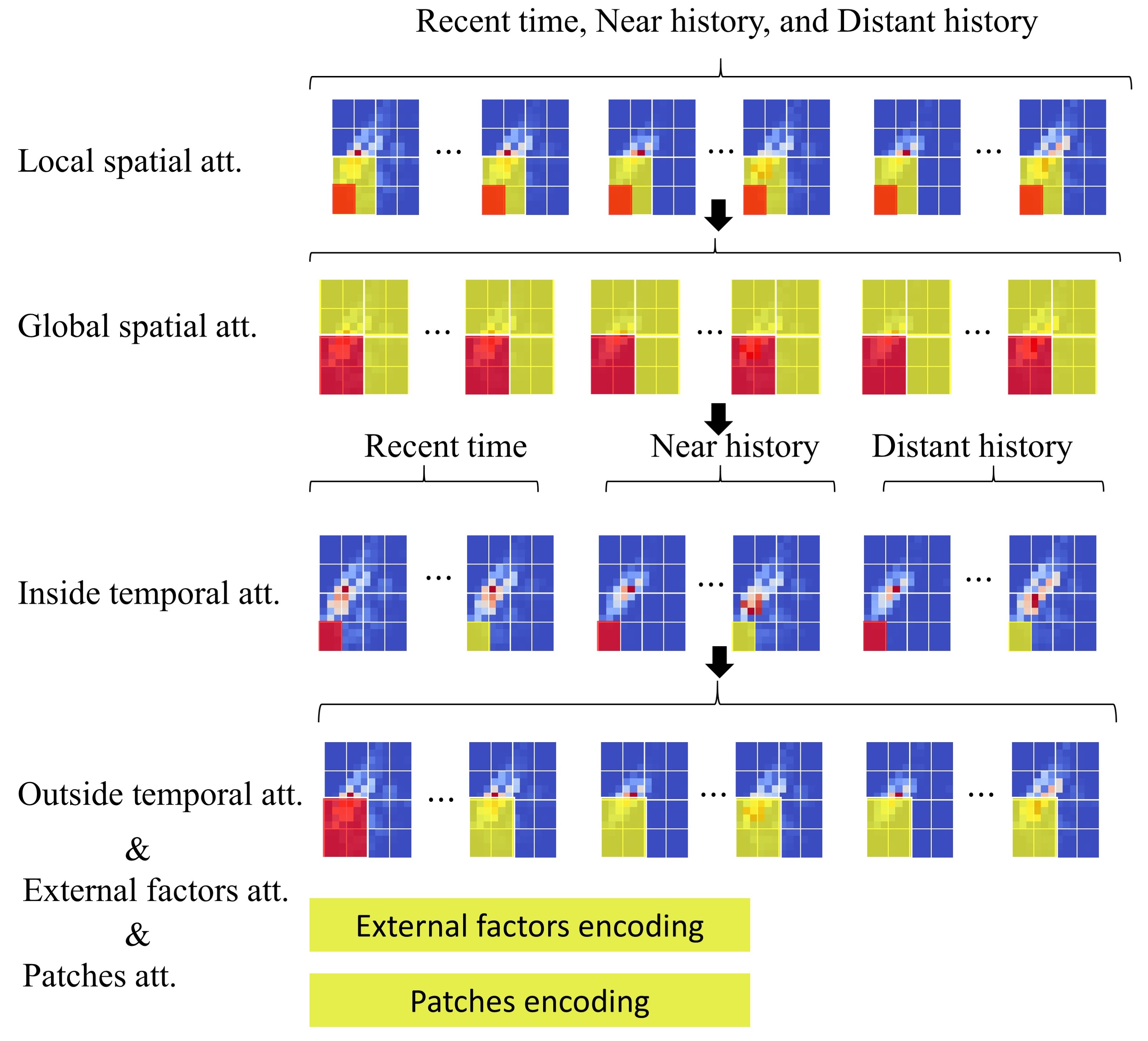}
    \caption{Visualization of the progressive space-time self-attention blocks.
    Each video is split to a sequence of  small patches with a size of 3 $\times$ 4 pixels.
    We show in red the query patch and show in yellow its spatiotemporal self-attention patches under each block.
    Patches without color are not used for the self-attention computation of the red patch.
    Note that self-attention is computed for every single patch in the video clip, i.e., every patch serves as a query.
    The self-attention pattern extends in the same fashion to all frames of the clip.
    The external factors and patches self-attention are computed with all patches in the outside temporal self-attention block.}
    \label{prostformer_image}
    \end{figure}

\begin{figure*}[ht]
    \centering
    \includegraphics[width=0.9\textwidth]{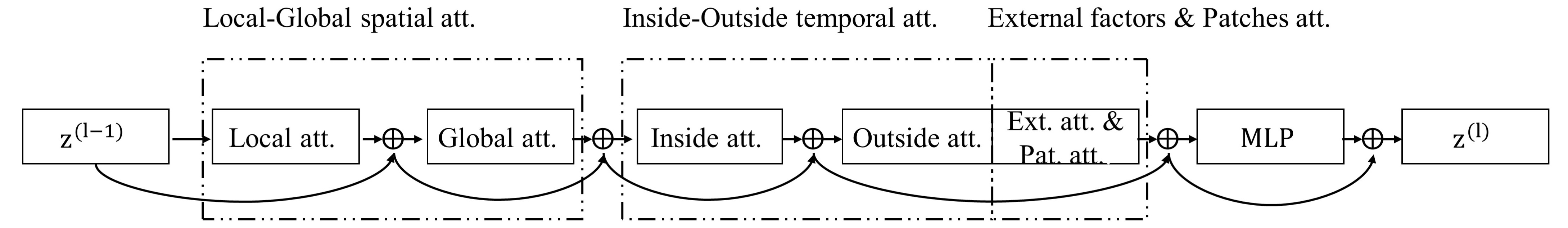}
    \caption{Illustration of the progressive self-attention blocks. We use residual connections to aggregate information from different self-attention layers within each block.
    A 1-hidden-layer MLP is applied after the outside temporal block.
    The final model is constructed by repeatedly stacking these blocks on top of each other.}
    \label{prostformer_eq}
    \end{figure*}

We first factorize the dependencies and then propose ProSTformer to focus on spatial dependence from local to global regions, on temporal dependence from inside to outside fragment (i.e., closeness, period, and trend), and finally on external dependence such as weather, temperature, and day-of-week.
In ProSTformer, each block highlights the unique dependence by aggregating the patches with spatiotemporal positions to significantly decrease the self-attention computation.
We give a sketch of the self-attention schemes in Figure
\ref{prostformer_image}, and provide architecture details in Figure \ref{prostformer_eq}.

\textbf{\emph{Local} spatial self-attention block.}
We split each frame into $N_1$ spatial groups, each of which  includes $N_2$ small patches.
Then, we implement self-attention for each small patch in the group, filtering the patches in other groups.
In practice, the group dimension $N_1$ and temporal dimension $F$ are merged into
the batch dimension for simplifying the operation. At the end of this block, we rearrange the output back to $(B, N F, D)$:
\begin{align}\label{pro_att_0}
    \operatorname{rearrange}(B, N F, D \to B N_1 F, N_2, D ).
\end{align}

For each small patch ${x}_{(p,t)}\in  \mathbb{R}^{2 \times P_2 \times
P_2^{\prime}}$  in each frame $t^{\prime}$, we only allow each key to attend its adjacent patches' queries:
\begin{equation}\label{space att}
    \bm{\alpha}_{(p, t^{\prime})}^{(l, a)}=\operatorname{SoftMax}\left(\frac{\mathbf{q}_{(p, t^{\prime})}^{(l, a)^{\top}}}{\sqrt{D_{h}}} \cdot \bm{k}_{(p^{\prime}, t^{\prime})}^{(l, a)} \right),
\end{equation}
where $p=1, \ldots, N $ and $p^{\prime}=1, \ldots, N_2$, note that $p^{\prime}$ is dynamically relative to $p$.

\textbf{\emph{Global} spatial self-attention block.}
After implementing the local spatial self-attention for each small patch,
we then implement global spatial self-attention for each big patch in the same frame.
In practice, the temporal dimension $F$ is merged into the batch dimension for simplifying the operation. At the end of this block, we rearrange the output back to $(B, N F, D)$:
\begin{equation}
    \operatorname{rearrange}(B, N F, D \to B F, N_1 , N_2 D ).
\end{equation}

For each big patch ${x}_{(p,t)}\in  \mathbb{R}^{2 \times P_1 \times P_1^{\prime}}$ in each frame $t^{\prime}$, we allow each key to attend the global patches' queries in Eq. \ref{space att},  where $p=1, \ldots, N_1 $ and $p^{\prime}=1, \ldots, N_1$.

\textbf{\emph{Inside} temporal self-attention block.}
We split the small patches in the same spatial position into $F_1$  temporal groups, each of which includes $F_2$ patches.
Then, we implement self-attention for each small patch inside the group, filtering the patches outside the group.
In practice, the group dimension $F_1$ and the spatial dimension $N$ are merged into the batch dimension for simplifying the operation. At the end of this block, we rearrange the output back to $(B, N F, D)$:
\begin{align}
    \operatorname{rearrange}(B, N F, D \to B N F_1, F_2, D ).
\end{align}

For each small patch ${x}_{(p,t)}\in  \mathbb{R}^{2 \times P_2 \times P_2^{\prime}}$  in the same spatial position $p^{\prime}$, we only allow each key to attend the inside patches' queries in the group:
\begin{equation}\label{time att}
    \bm{\alpha}_{(p^{\prime}, t)}^{(l, a)}=\operatorname{SoftMax}\left(\frac{\mathbf{q}_{(p^{\prime}, t)}^{(l, a)^{\top}}}{\sqrt{D_{h}}} \cdot \bm{k}_{(p^{\prime}, t^{\prime})}^{(l, a)} \right),
\end{equation}
where $t=1, \ldots, F $ and $t^{\prime}=1, \ldots, F_2$, note that $t^{\prime}$ is dynamically relative to $t$.

\textbf{\emph{Outside} temporal self-attention block.}
After implementing the inside temporal self-attention,
we then implement the outside temporal self-attention for each big patch in the same spatial position. In practice, the spatial dimension $N_1$ is merged into the batch dimension for simplifying the operation, at the end of this block, we rearrange the output back to $(B, N F, D)$:
\begin{equation}\label{pro_att_1}
    \operatorname{rearrange}(B, N F, D \to B N_1, F , N_2 D ).
\end{equation}

For each big patch ${x}_{(p,t)}\in  \mathbb{R}^{2 \times P_1 \times P_1^{\prime}}$ in same spatial position $p^{\prime}$,  we allow each key to attend the patches' queries of all frames in Eq. \ref{time att}, where $t=1, \ldots, F $ and $t^{\prime}=1, \ldots, F$.


\textbf{Encoding.} We encode the values of patches in every block. The encoding  $\mathbf{z}_{(p, t)}^{(l)}$  at block $l$  are obtained by weighting the sum of value vectors using self-attention coefficients from each self-attention head:
\begin{equation}
    \bm{s}_{(p, t)}^{(l, a)}=\sum_{p^{\prime}=1}^{N} \sum_{t^{\prime}=1}^{F} \bm{\alpha}_{(p, t),\left(p^{\prime}, t^{\prime} \right)}^{(l, a)} \mathbf{v}_{\left(p^{\prime}, t^{\prime}\right)}^{(l, a)}.
\end{equation}

Note that we implement spatial self-attention only along the spatial dimension, and the temporal self-attention only  along the temporal dimension. $p,p^{\prime},N,t,t^{\prime},F $ have different  definitions in each block.
Then, we concatenate these vectors from all heads and linearly map it back to patch dimension, using residual connections after LayerNorm:
\begin{align}
    \mathbf{z^{\prime}}_{(p, t)}^{(l)}&=W[\mathbf{s}_{(p, t)}^{(l, a)}, \ldots ,\mathbf{s}_{(p, t)}^{(l, A)}   ]^{\top} \\
    \mathbf{z}_{(p, t)}^{(l)}&=\operatorname{LN}(\mathbf{z^{\prime}}_{(p, t)}^{(l)})+\mathbf{z}_{(p, t)}^{(l-1)} . \label{encoding}
\end{align}

The encoding process is conducted in each self-attention block of ProSTformer.
For simplicity, we do not replicate it in each block.
In the final outside temporal self-attention block, the output is passed to an MLP layer.

\textbf{External factors embedding.} We linearly map the external factors such as weather, day-of-week, wind speed and temperature into an embedding token [ext] $\mathbf{z}_{ext}^{(l)} \in  \mathbb{R}^{N_2D}$, then compute  self-attention and encoding with all patches in the  outside temporal self-attention block.
The output is passed to an MLP layer.

\textbf{Patches embedding.} We extend the concept of token [cls], using $N_1$ randomly initialized patches token [pat] $\mathbf{z}_{pat}^{(N_1,l)} \in  \mathbb{R}^{N_2D}$ to compute self-attention and encoding with all patches in the  outside temporal self-attention block. The final prediction $\mathbf{Pre} \in  \mathbb{R}^{2\times H\times W}$ is obtained by combining the [pat] tokens:
\begin{equation}
    \mathbf{Pre}=\operatorname{combine}[\mathbf{z}_{pat}^{(n,l)},\ldots , \mathbf{z}_{pat}^{(N_1,l)}].
\end{equation}

\subsection{ProSTformer Model}
We reduce the self-attention computation by replacing with the progressive space-time self-attention in Eq. \eqref{pro_att_0} to \eqref{pro_att_1}.
For each patch, excluding the [ext] and [pat] self-attention, compared to $(NF)$ times self-attention needed by Vanilla  Transformer, ProSTformer only need $(N_2+N_1+F_2+F)$ times self-attention.
The encoding $\mathbf{z}_{(p, t)}^{(l)}$ resulting from  local spatial self-attention block is then fed to global spatial self-attention block, temporal self-attention block, and outside temporal self-attention block instead of being passed to the MLP.
Finally, the encoding of the outside temporal self-attention is passed to the MLP.
In each block, new key/query/value vectors are obtained from Eq. \eqref{qkv} and matrices $\left\{W_{Q^{\text {local }}}^{(l, a)}, W_{K^{\text {local }}}^{(l, a)}, W_{V^{\text {local }}}^{(l, a)}\right\}$,
 $\left\{W_{Q^{\text {global }}}^{(l, a)}, W_{K^{\text {global}}}^{(l, a)}, W_{V^{\text {global}}}^{(l, a)}\right\}$,
 $\left\{W_{Q^{\text {inside }}}^{(l, a)}, W_{K^{\text {inside}}}^{(l, a)}, W_{V^{\text {inside }}}^{(l, a)}\right\}$,
 and $\left\{W_{Q^{\text {outside }}}^{(l, a)}, W_{K^{\text {outside}}}^{(l, a)}, W_{V^{\text {outside}}}^{(l, a)}\right\}$.
To explore the time and space priority for traffic flow forecasting, 
we also experiment with a progressive ``time-space" self-attention model ProTSformer that reverses the time and space priority of ProSTformer.

\section{Experiment}
\begin{table*}[t]
    \caption{The prediction results. ProST$^{\dagger}$ is pre-trained on  30 min datasets and then fine tuned on  60 min and 90 min datasets. Other methods have no pre-training.}
    \label{table 1}
    \centering
    \begin{tabular}{|c|c|c|c|c|c|c|c|c|c|c|}
    \hline
    \multirow{7}{*}{\begin{tabular}[c]{@{}c@{}}NYC\\ Taxi\end{tabular}} &
      Win. &
      Metrix &
      HA &
      ConvLSTM &
      ST-Res &
      Trans &
      Informer &
      TimeS &
      ProST &
      ProST$^{\dagger}$ \\ \cline{2-11} 
     & \multirow{2}{*}{90 min} & RMSE & 119.32 & 26.43         & 15.56         & 20.97 & 15.98 & 20.38 & 16.71           & \textbf{12.79}   \\ \cline{3-11} 
     &                         & MAE  & 32.83  & 7.58          & 4.91          & 6.85  & 5.40  & 6.82  & 5.55            & \textbf{4.56}    \\ \cline{2-11} 
     & \multirow{2}{*}{60 min} & RMSE & 80.58  & 34.20         & 11.32         & 14.98 & 11.72 & 15.09 & 11.71           & \textbf{10.54}   \\ \cline{3-11} 
     &                         & MAE  & 22.26  & 8.49          & 4.07          & 5.57  & 4.45  & 5.63  & 4.39            & \textbf{4.01}    \\ \cline{2-11} 
     &
      \multirow{2}{*}{30 min} &
      RMSE &
      40.92 &
      22.14 &
      10.52 &
      13.50 &
      10.30 &
      13.19 &
      \multicolumn{2}{c|}{\textbf{10.21}} \\ \cline{3-11} 
     &                         & MAE  & 11.45  & 6.33          & \textbf{3.64} & 5.10  & 3.98  & 4.95  & \multicolumn{2}{c|}{3.82}          \\ \hline
    \multirow{6}{*}{\begin{tabular}[c]{@{}c@{}}NYC\\ Bike\end{tabular}} &
      \multirow{2}{*}{90 min} &
      RMSE &
      25.62 &
      1.99 &
      2.10 &
      2.09 &
      2.02 &
      2.06 &
      1.99 &
      \textbf{1.91} \\ \cline{3-11} 
     &                         & MAE  & 8.47   & \textbf{0.78}          & 0.85          & 0.84  & 0.93  & 0.86  & \textbf{0.78}   & 0.79             \\ \cline{2-11} 
     & \multirow{2}{*}{60 min} & RMSE & 17.43  & 4.59          & 4.06          & 4.48  & 4.13  & 4.44  & 3.97            & \textbf{3.78}    \\ \cline{3-11} 
     &                         & MAE  & 5.76   & 1.66          & \textbf{1.61}          & 1.83  & 1.72  & 1.80  & 1.63            & \textbf{1.61}    \\ \cline{2-11} 
     & \multirow{2}{*}{30 min} & RMSE & 8.90   & \textbf{2.30}          & 2.47          & 2.45  & 2.34  & 2.43  & \multicolumn{2}{c|}{\textbf{2.30}} \\ \cline{3-11} 
     &                         & MAE  & 3.00   & \textbf{1.01} & 1.09          & 1.10  & 1.06  & 1.09  & \multicolumn{2}{c|}{1.04}          \\ \hline
    \multicolumn{3}{|c|}{Count}       & 0      & 3             & 2             & 0     & 0     & 0     & 3               & 9                \\ \hline
    \end{tabular}
    \end{table*}

\subsection{Datasets} We use two datasets, including the  trajectory data of yellow taxi and sharing bike in New York City (NYC).

\textbf{NYCTaxi:} The trajectory data is yellow taxi GPS data for New York City (NYC) from 1st Jan. 2013 to 31th Dec. 2015 about 416 million trajectories.  We  partition NYC into $12 \times 16$ regions.  To explore the dataset scale granularity in this problem, we create separate datasets as $\{\operatorname{NYCTaxi}_{\operatorname{30min}}, \operatorname{NYCTaxi}_{\operatorname{60min}}, \operatorname{NYCTaxi}_{\operatorname{90min}} \}$ for 30 minutes, 60 minutes, and 90 minutes time windows.

\textbf{NYCBike:} The trajectory data is sharing bike GPS data for New York City (NYC) from 1st Jan. 2018 to 31th Dec. 2020, about 56 million trajectories.
We partition NYC into $12 \times 16$ regions, and create separate datasets as
$\{\operatorname{NYCBike}_{\operatorname{30min}}, \operatorname{NYCBike}_{\operatorname{60min}}, \operatorname{NYCBike}_{\operatorname{90min}} \}$.

For all datasets, we choose data from the last four weeks as the test set, all data before that as the training set. We remove unavailable bike stations in NYCBike, and the city regions' partition is same in NYCTaxi and NYCBike.

\subsection{Experimental Details}
\textbf{Baselines}: We select six  forecasting methods as the comparison.
For Transformer, TimeSformer, and Informer, we implement  the same external factors and patches self-attention with ProSTformer.
We split per frame into 16 patches for 12 frames, totally obtain 192 patches as input.
All methods have the same input, except that ConvLSTM and
HA exclude the external factors.
\begin{itemize}
    \item \textbf{HA :} Historical average,  uses the average of previous values in the training dataset as the prediction.
    \item \textbf{ConvLSTM \cite{xingjian2015convolutional} :}  ConvLSTM adds a convolutional structure to LSTM to  learn spatiotemporal features.
    \item \textbf{ST-ResNet \cite{zhang2017deep} :} ST-ResNet employs convolution-based residual networks to extract spatiotemporal dependence.
    \item \textbf{Transformer \cite{dosovitskiy2020image} :} Transformer directly learns the spatiotemporal features from a sequence of video patches' embedding.
    \item \textbf{Informer \cite{zhou2021informer}} Informer designs a ProbSparse self-attention mechanism distilling highlights dominating self-attention by halving cascading layer input.
    \item \textbf{TimeSformer \cite{bertasius2021space}:} TimeSformer proposes a ``divided self-attention", where the temporal and spatial self-attention are separately applied within each block.
\end{itemize}

\textbf{Hyper-parameter tuning:} ProSTformer contains 6-layer stack. Similar to the setting in \cite{dosovitskiy2020image}, our method is optimized with Adam optimizer for 1000 epochs, and its learning rate linearly warmup from 0 to 1e-4 for 200 epochs and then linearly decreases to 0. The batch size is 32 per GPU. The details of baseline methods can be found in Appendix A. \textbf{Normalization:} The input of each dataset is Min-Max normalized to the range [-1,1].  \textbf{Metrics:} We train all methods with MSE loss, and use two evaluation metrics: $\operatorname{RMSE}=\sqrt{ \frac{1}{n} \sum_{i=1}^{n}(\mathbf{y}-\hat{\mathbf{y}})^{2} }$ and $\operatorname{MAE}=\frac{1}{n} \sum_{i=1}^{n}|\mathbf{y}-\hat{\mathbf{y}}|$. \textbf{Platform:} All the methods were trained/tested on double Nvidia V100 16GB GPUs.

\subsection{Results and Analysis}
\begin{table}[ht]
    \caption{The performances of data augmentation tasks on  30 min NYCTaxi dataset. ProST is the benchmark without pre-training.}
    \label{augmentation}
    \centering
    \begin{tabular}{|c|c|l|l|l|}
    \hline
    Tasks & ProST & \multicolumn{1}{c|}{Rotation} & \multicolumn{1}{c|}{Time-point} \\ \hline
    RMSE  & 10.21     & 10.60                     & 10.19                                              \\ \hline
    MAE   & 3.82      & 3.99                     & 3.82                                                \\ \hline
    \end{tabular}
    \end{table}

\begin{table*}[ht]
    \caption{The prediction results with pre-training. All methods are pre-trained on 30 min datasets, and then fine tuned on  60 min and the 90 min  datasets. The "-s" indicates the input length is 48 rather than 196 for other models. ProTS$^{\dagger }$ reverses the spatial and temporal blocks of ProST$^{\dagger }$.}
    \label{pre-training}
    \centering
\begin{tabular}{|c|c|c|c|c|c|c|c|c|c|c|c|}
\hline
\multirow{7}{*}{\begin{tabular}[c]{@{}c@{}}NYC\\ Taxi\end{tabular}} &
  Win. &
  Metrix &
  ConvLSTM &
  ST-Res &
  Trans &
  Trans-s &
  Informer &
  TimeS &
  TimeS-s &
  ProTS$^{\dagger }$ &
  ProST$^{\dagger }$ \\ \cline{2-12} 
 & \multirow{2}{*}{90 min} & RMSE & 24.76         & 13.89         & 18.12 & 13.26 & \textbf{12.67} & 17.71 & 13.34         & 13.51         & 12.79          \\ \cline{3-12} 
 &                         & MAE  & 6.92          & 4.79          & 6.11  & 4.70  & 4.69           & 6.09  & 4.76          & 4.84          & \textbf{4.56}  \\ \cline{2-12} 
 & \multirow{2}{*}{60 min} & RMSE & 33.41         & 11.26         & 14.61 & 10.91 & 10.77          & 14.24 & 10.78         & 10.92         & \textbf{10.54} \\ \cline{3-12} 
 &                         & MAE  & 7.96          & 4.14          & 5.44  & 4.14  & 4.35           & 5.40  & 4.13          & 4.22          & \textbf{4.01}  \\ \cline{2-12} 
 & \multirow{2}{*}{30 min} & RMSE & 22.14         & 10.52         & 13.50 & 10.43 & 10.30          & 13.19 & 10.29         & 10.64         & \textbf{10.21} \\ \cline{3-12} 
 &                         & MAE  & 6.33          & \textbf{3.64} & 5.10  & 3.89  & 3.98           & 4.95  & 3.89          & 4.01          & 3.82           \\ \hline
\multirow{6}{*}{\begin{tabular}[c]{@{}c@{}}NYC\\ Bike\end{tabular}} &
  \multirow{2}{*}{90 min} &
  RMSE &
  1.92 &
  1.98 &
  2.04 &
  \textbf{1.89} &
  1.91 &
  2.03 &
  \textbf{1.89} &
  1.93 &
  1.91 \\ \cline{3-12} 
 &                         & MAE  & \textbf{0.76} & 0.80          & 0.83  & 0.78  & 0.83           & 0.82  & 0.77          & 0.78          & 0.79           \\ \cline{2-12} 
 & \multirow{2}{*}{60 min} & RMSE & 4.51          & 4.12          & 4.33  & 3.74  & 3.76           & 4.30  & 3.72          & \textbf{3.71} & 3.78           \\ \cline{3-12} 
 &                         & MAE  & 1.62          & 1.65          & 1.76  & 1.58  & 1.65           & 1.75  & \textbf{1.56} & 1.58          & 1.61           \\ \cline{2-12} 
 & \multirow{2}{*}{30 min} & RMSE & \textbf{2.30}          & 2.47          & 2.45  & \textbf{2.30}  & 2.34           & 2.43  & \textbf{2.30} & 2.32          & \textbf{2.30}  \\ \cline{3-12} 
 &                         & MAE  & \textbf{1.01} & 1.09          & 1.10  & 1.03  & 1.06           & 1.09  & 1.03          & 1.04          & 1.04           \\ \hline
\multicolumn{3}{|c|}{Count}       & 3             & 1             & \multicolumn{2}{c|}{2}        & 1              &\multicolumn{2}{c|}{3}             & \multicolumn{2}{c|}{6}         \\ \hline
\end{tabular}
\end{table*}
\begin{table*}[t]
    \caption{Ablation study of each block. ProST is the benchmark of no ablation.}
    \label{Ablation}
    \centering
    \begin{tabular}{|c|c|c|c|c|c|}
    \hline
    Block ablation  & ProST    & No the local block & No the global  block & No  the Inside  block & No the Outside  block\\ \hline
    RMSE & 10.21 & 10.24      & 10.25        & 10.24        & 10.36         \\ \hline
    MAE  & 3.82  & 3.88       & 3.80        & 3.86         & 3.84          \\ \hline
    \end{tabular}
    \end{table*}

Table \ref{table 1} summarizes the evaluation results of all  methods on two datasets. We gradually prolong the time window on NYCTaxi and NYCBike, a longer time window means a smaller dataset scale, a 30 min dataset contains 52,560 items, a 60 min dataset contains 26,280 items, and a 90 min dataset contains 17,520 items.
And we define 30 min datasets as big scale,  60 min datasets as medium scale, and  90 min datasets as small scale.

\begin{figure}[t]
    \begin{minipage}[t]{0.49\linewidth}
      \centering
      \includegraphics[scale=0.29]{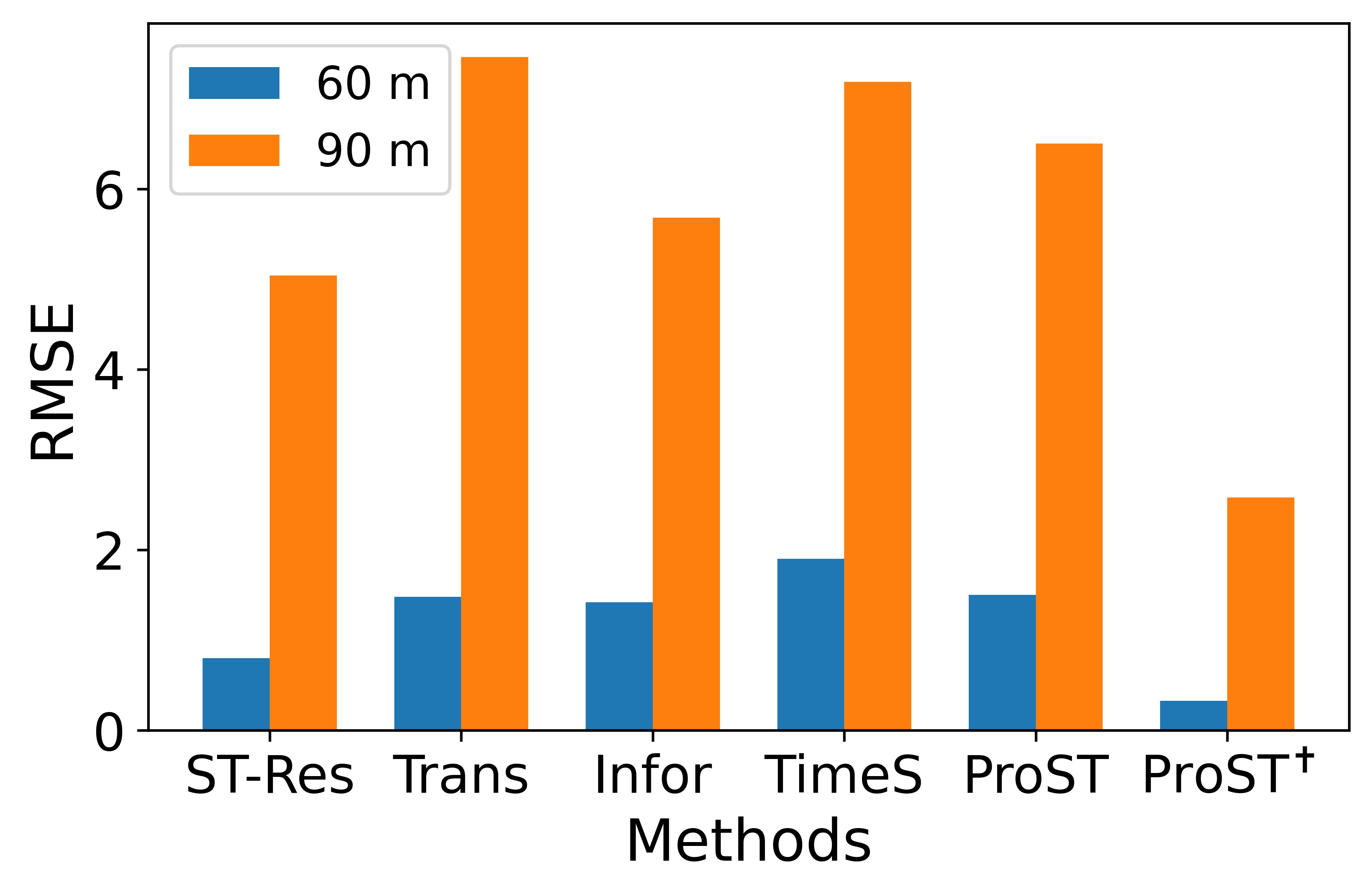}
      \caption{The error increase on   NYCTaxi datasets.}
      \label{increase}
    \end{minipage}\hspace{1 mm}
    \begin{minipage}[t]{0.49\linewidth}
      \centering
      \includegraphics[scale=0.29]{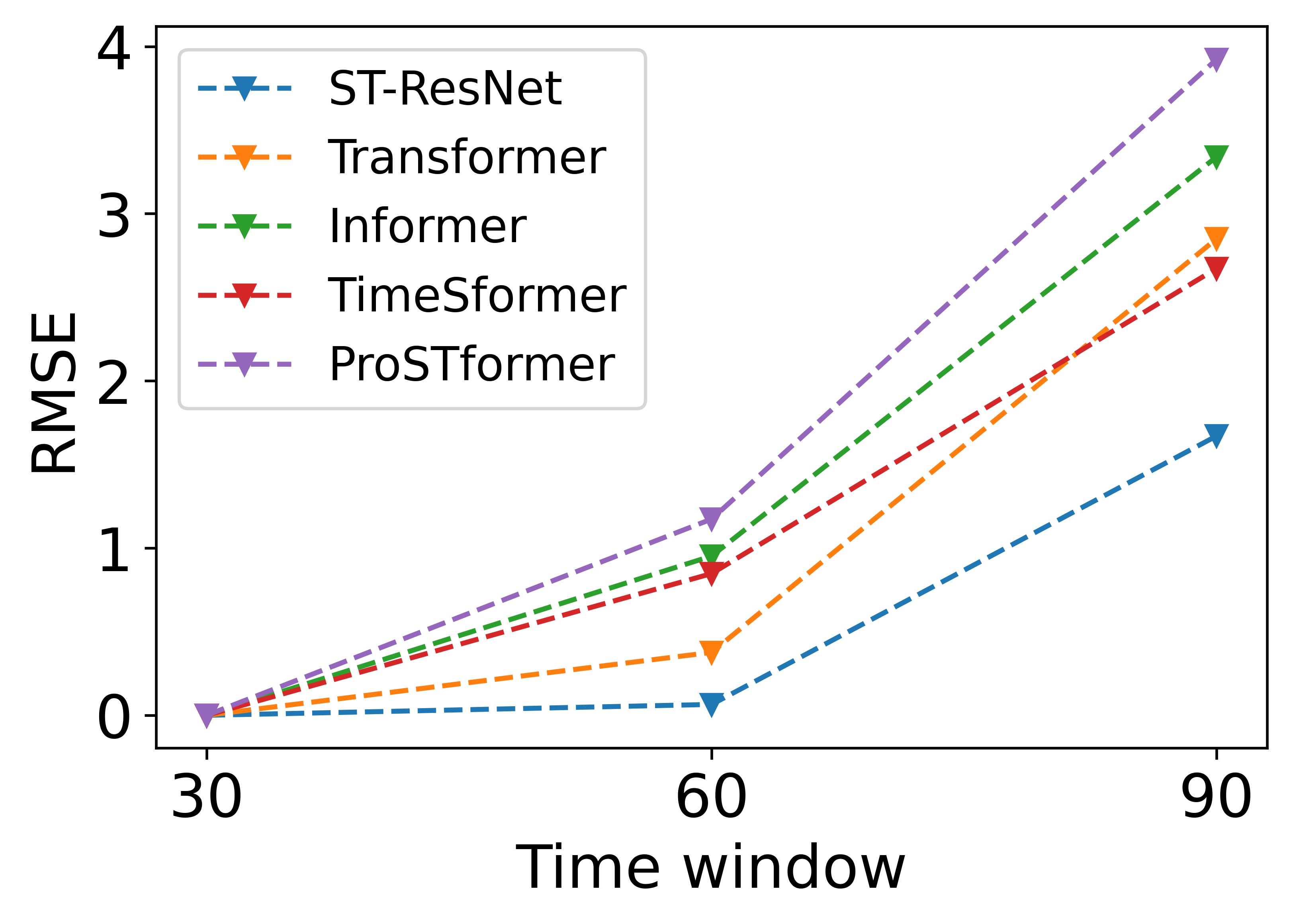}
      \caption{The error decrease  with pre-training on NYCTaxi datasets.}
      \label{decrease}
    \end{minipage}
  \end{figure}

\textbf{Forecasting on different time windows and dataset scales.}
From Table \ref{table 1}, we can observe that:
\textbf{(1)} The proposed  model ProSTformer$^{\dagger }$ that is pre-trained on 30 min datasets, significantly improves the prediction performance (wining-counts in the last column), and its error rises more slowly than other methods with the growing of dataset scale, as shown in Figure \ref{increase}.
\textbf{(2)} Without pre-training, ProSTformer performs better than the relevant TimeSformer, Informer, and Transformer on big scale datasets.
\textbf{(3)} Compared with ST-ResNet and ConvLSTM, ProSTformer performs better or the same on big scale datasets by RMSE. On medium and small scale datasets,  pre-trained model ProSTformer$^{\dagger }$ performs better than them by RMSE.
Specifically, compared with ST-ResNet by RMSE. On NYCTaxi, ProSTformer$^{\dagger }$ achieves a decrease 3.1\% on the big scale dataset, 7.5\% on the medium scale dataset, 21.7\% on the small scale dataset. On NYCBike, ProSTformer$^{\dagger }$ achieves a decrease of 7.0\% on the big scale dataset, 7.0\% on the medium scale dataset, 9.2\% on the small scale dataset.

\textbf{The importance of pre-training.}
Without pre-training, the errors of ProSTformer, TimeSformer, Informer, and
Transformer increase more rapidly than ST-ResNet on  medium and small scale NYCTaxi datasets, as shown in Figure \ref{increase}. 
But with pre-training,  ProSTformer  achieves a significant enhancement, as shown in Figure \ref{decrease},  and the error increase of ProSTformer$^{\dagger }$ is the slowest on the medium and small scale NYCTaxi datasets as shown in Figure \ref{increase}.

\begin{figure}[ht]
\centering
\includegraphics[width=0.4\textwidth]{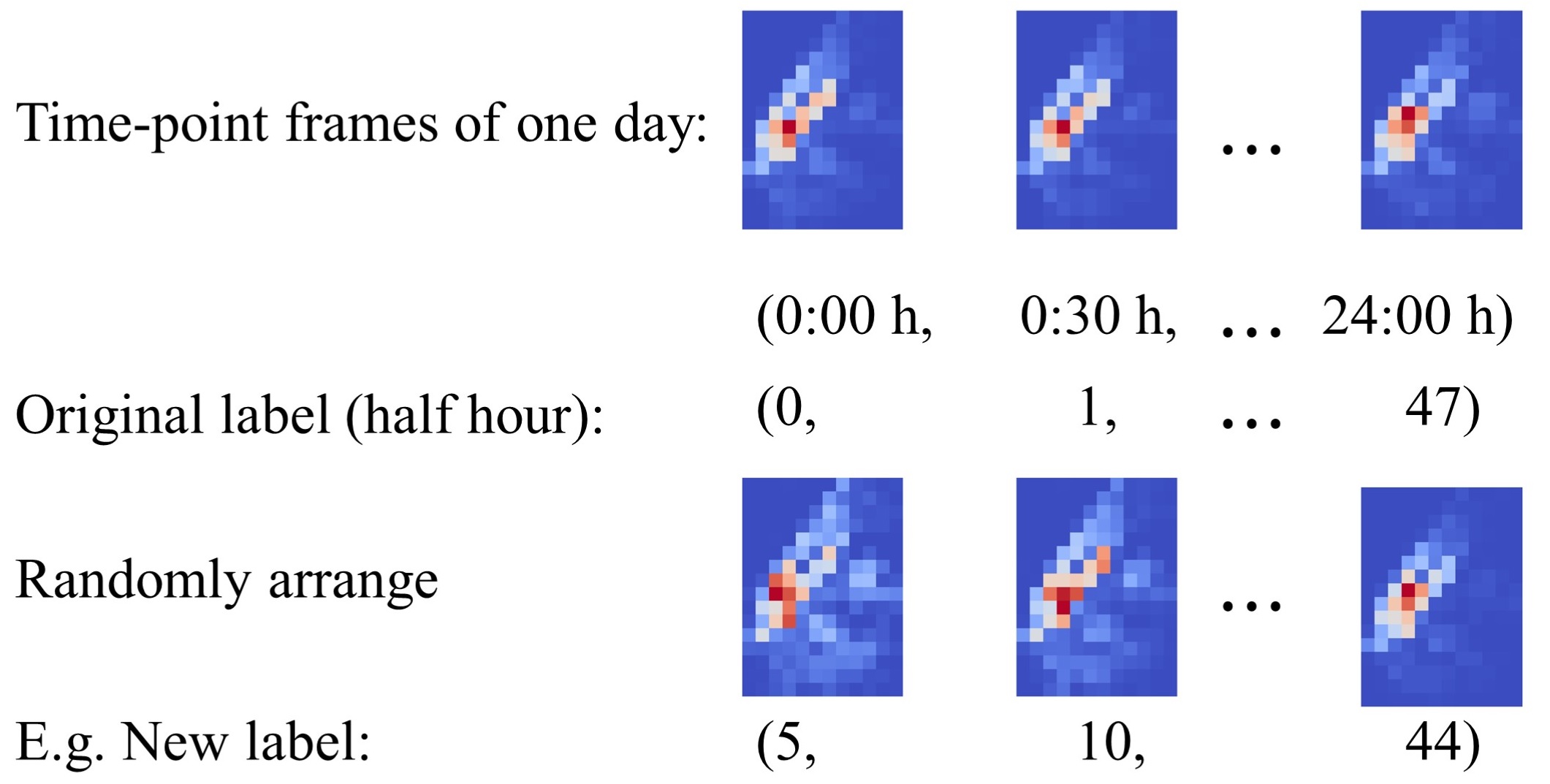}
\caption{The random arrangement of time-point frames.}
\label{time_point}
\end{figure}

\textbf{The performances of data augmentation tasks.}
We directly pre-train ProSTformer on 30 min datasets, and fine tune it on the 60 min and 90 min datasets, leading to the best performance model ProSTformer$^{\dagger }$.
From the perspective of data augmentation, the pre-training on  30 min datasets can be regarded as a closely relevant augmentation task for traffic flow forecasting task on 60 min and 90min datasets.
We also conduct additional two data augmentation tasks to pre-train ProSTformer, which are image rotation class task \cite{gidaris2018unsupervised} and image time-point class task, as shown in Figure \ref{time_point}, based on  30 min NYCTaxi dataset.
From Table \ref{augmentation}, we observe that the rotation class task increases the error, and the time-point class task slightly decreases the error.
The two tasks do not improve the performances possibly because the two tasks are low relevant to traffic flow forecasting task for extracting the spatiotemporal dependence.


\subsection{Fine tune the baseline methods}

On account of  pre-training  significantly improve the performance of ProSTformer,  we also implement same pre-training with all the baseline methods.

\textbf{The effects of pre-training.} 
We observe that: \textbf{(1)}
From Table \ref{pre-training}, on a total of six datasets,
our method obtains top 1 on four datasets and top 2 on the other two datasets by RMSE.
On the whole, our method performs best (wining-counts in the last column), and the enhancement of ProSTformer$^{\dagger }$ is the most significant with pre-training on NYCTaxi, as shown in Figure \ref{decrease}.
\textbf{(2)}
Compare Table \ref{table 1} with Table \ref{pre-training},  pre-training almost improves all methods' performances.
With pre-training and shorter input length, Informer, Transformer-s, and TimeSformer-s instead surpass ST-ResNet by RMSE.
It shows the convolution-free Transformer-like model’s potential value  in traffic flow forecasting. 

\textbf{The priority of space and time.} 
We explore the time and space priority in ProSTformer$^{\dagger }$ on different datasets.
From Table \ref{pre-training},  the space priority model ProSTformer$^{\dagger }$ generally performs better (wining-counts in the last column) than the time priority model ProTSformer$^{\dagger }$, 
but ProTSformer$^{\dagger }$ performs better on 60 min NYCBike dataset by RMSE. 


\textbf{The performances of different sequence length.}
We also conduct additional experiments with 48 sequence length  for Transformer and TimeSformer rather than 196, which is denoted as ``Trans-s" and ``TimeS-s" in Table \ref{pre-training}.
It is shown that the shorter input sequence significantly improves the performances of Transformer and TimeSformer.
The results verify the assumption that there are redundant dot-product pairs of the low relevant patches.
ProSTformer$^{\dagger}$ performs well on the long input sequence, 
we attribute it to that the progressive self-attention mechanism significantly decrease the redundant computation.
Informer also performs well on the long input sequence, because its distilling mechanism effectively filters the redundant computation.

\subsection{Ablation Study}
We also conducted additional experiments on  30 min NYCTaxi dataset with ablation consideration for ProSTformer.

\textbf{The performance of each block.}
We explore the influence of each block in ProSTformer by
eliminating the local, global, inside, and outside self-attention block, respectively.
The other experimental setups are aligned with the settings of ProSTformer. From Table \ref{Ablation}, ProSTformer achieves the best performance than the ablation version models by RMSE.
We conclude that the progressive space-time self-attention mechanism is worth adopting.

\textbf{The performances of decomposition ways.}
\begin{figure}[t]
    \centering
    \subfigure[way 1] {
        \label{way 1}
    \includegraphics[scale=0.05]{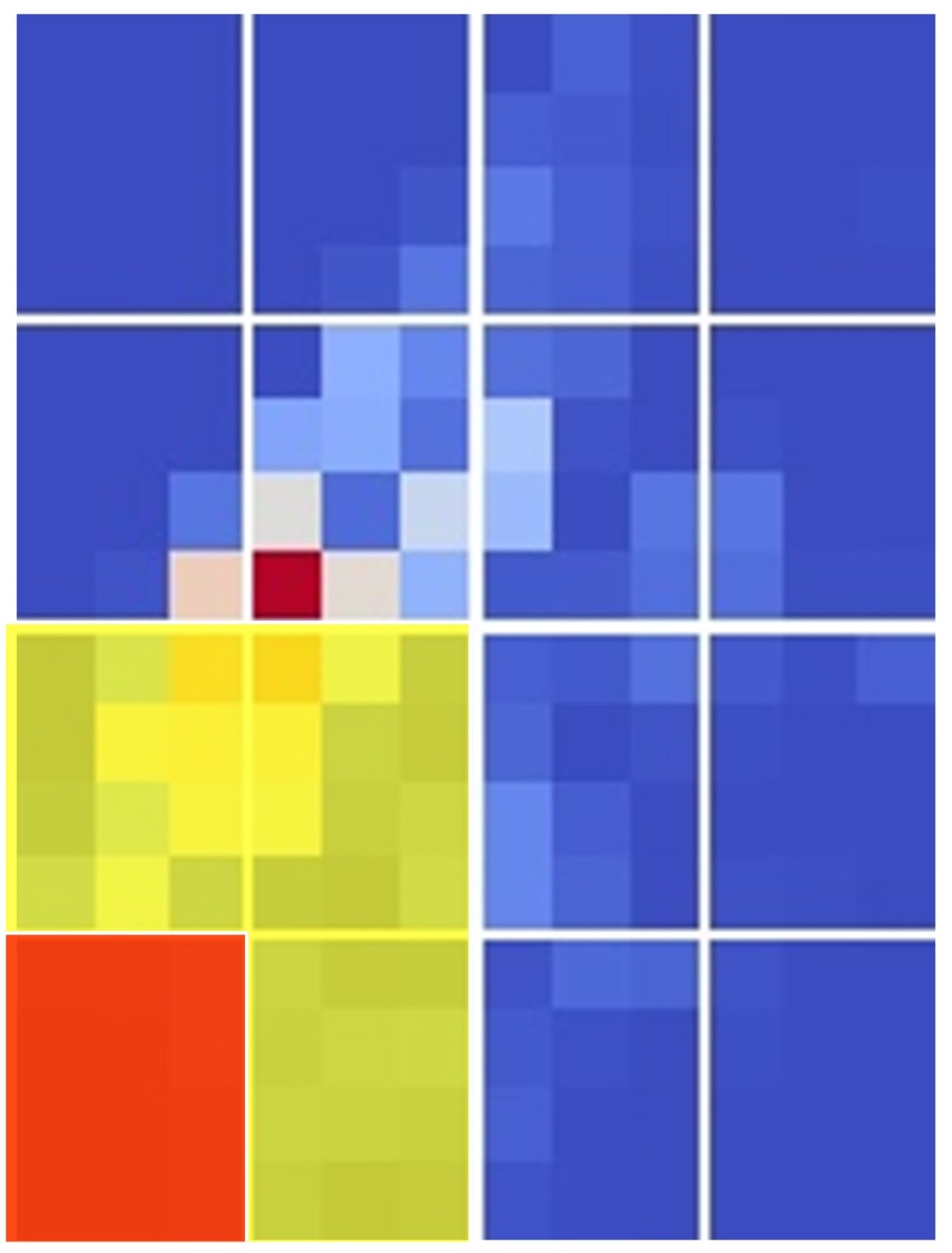}
    }
    \subfigure[way 2] {
    \label{way 2}  \hspace{1 mm}
    \includegraphics[scale=0.05]{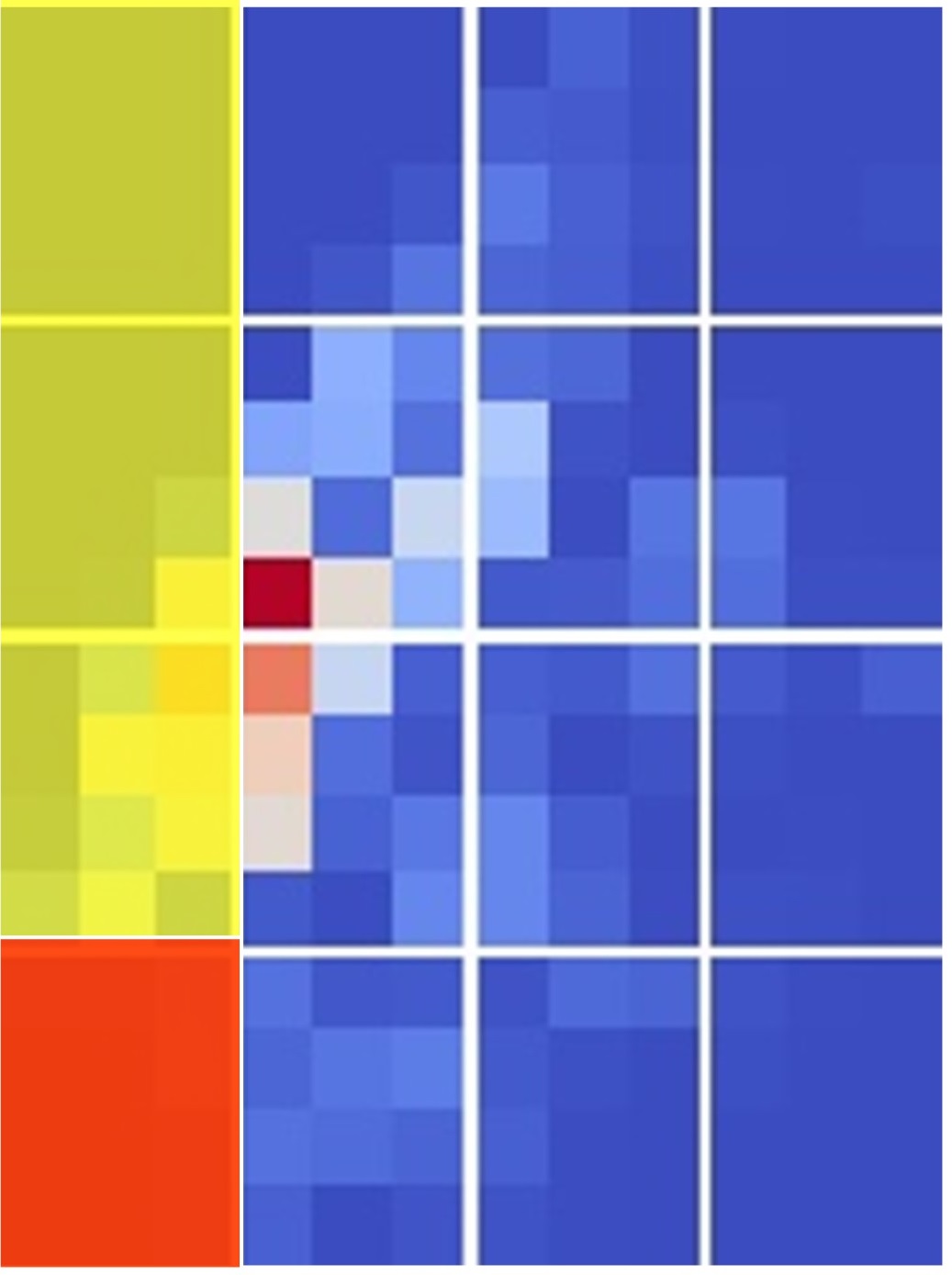}      }
    \subfigure[way 3] {
        \label{way 3}     \hspace{1 mm}
    \includegraphics[scale=0.05]{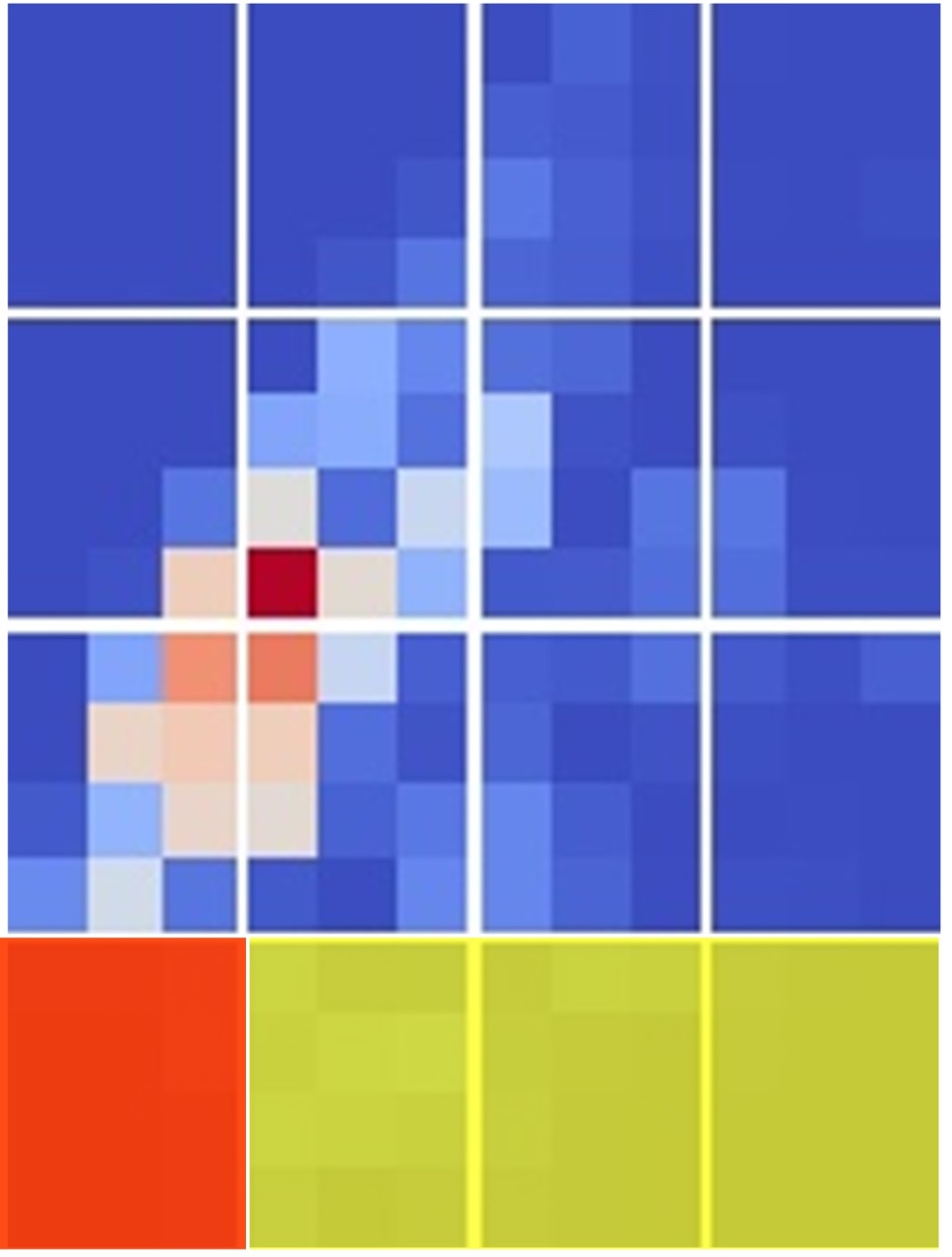}   }
    \caption{The decomposition ways. We show in red the
    query patch and show in yellow its local spatial self-attention
    patches.}
    \label{decomposition ways}
    \end{figure}
\begin{table}[t]
    \caption{Ablation study of decomposition ways.}
    \label{decomposition}
    \centering
    \begin{tabular}{|c|c|c|l|c|}
    \hline
    \multicolumn{1}{|l|}{Decomposition ways} & Way 1   & Way 2 & Way 3                    \\ \hline
    RMSE                       & 10.21 & 10.35   & \multicolumn{1}{c|}{10.30}       \\ \hline
    MAE                        & 3.82  & 3.86    & 3.82                            \\ \hline
    \end{tabular}
    \end{table}
In Table \ref{table 1}, \ref{augmentation}, \ref{pre-training},  and \ref{Ablation}, we use way 1 in Figure \ref{way 1} to decompose the flow videos and implement the self-attention in ProSTformer.
The decomposition way indicates the adjacent patches of a patch to implement  the local-global spatial self-attention.
The spatial self-attention blocks will influence the following temporal blocks due to the subsequent encoding pass.
We conduct additional experiments with decomposition way 2 and way 3, as shown in Figure \ref{way 2} and \ref{way 3}, to explore the decomposition ways influence.
From Table \ref{decomposition}, way 1 intuitively and practically performs better than  way 2 and way 3 by RMSE.

\section{Conclusion}
In this paper, we propose  ProSTformer for traffic flow forecasting.
We first factorize the dependencies and then design a corresponding progressive space-time self-attention mechanism to extract the dependencies.
The mechanism  incorporates the structure information of the flow videos to significantly decrease the self-attention computation.
We show that pre-training is of great importance on  medium scale and  small scale datasets for Transformer-like models.
The experiments on real-world datasets demonstrate that ProSTformer enhances the prediction capacity significantly.

\bibliography{intsformer}

\end{document}